\documentclass[traditabstract,letter]{aa}

\usepackage{natbib}
\usepackage{epsfig}
\usepackage{epsf}
\usepackage{color}
\definecolor{red}{rgb}{0.7,0,0}
\definecolor{blue}{rgb}{0,0,0.7}
\def\correc#1{#1}
\def\corr#1{#1}

\usepackage{amssymb}

\def\ergcms{erg~cm$^{-2}$~s$^{-1}$}
\def\nh{N$_\mathrm{H}$}
\def\cm2{cm$^{-2}$}

\def\integral{{\it{INTEGRAL}}}
\def\swift{{\it{Swift}}}

\def\xte{{\it{RXTE}}}
\def\chisq{$\chi^2_\nu$}

\def\IGR{IGR~J17091$-$3624}
\def\atca{{\it{ATCA}}}

\usepackage{longtable}
\usepackage{natbib}
 \usepackage[figuresright]{rotating}
\usepackage{lscape}
\usepackage{txfonts}
\begin{document}
   \title{First simultaneous multi-wavelength observations of the black hole candidate \IGR }
\subtitle{\atca, \integral, \swift, and \xte\ views of the 2011 outburst}

   \author{J. Rodriguez\inst{1} \and 
           S. Corbel\inst{1} \and I. Caballero\inst{1}   \and J.A. Tomsick\inst{2} \and T. Tzioumis\inst{3} \and A. Paizis\inst{4} \and M. Cadolle Bel\inst{5} \and E. Kuulkers\inst{5} }

   \offprints{J. Rodriguez:  jrodriguez@cea.fr}
\authorrunning{Rodriguez  et al. }

\titlerunning{Multi-wavelength observations of \IGR }
 
   \institute{Laboratoire AIM, CEA/IRFU - CNRS/INSU - Universit\'e Paris Diderot, CEA DSM/IRFU/SAp, F-91191 Gif-sur-Yvette, France
  \and  Space Sciences Laboratory, 7 Gauss Way, University of California, Berkeley, CA 94720-7450, USA
   \and Australia Telescope National Facility, CSIRO, P.O. Box 76, Epping NSW 1710, Australia
 \and Istituto Nazionale di Astrofisica, INAF-IASF, Via Bassini 15, 20133 Milano, Italy
   \and ESAC, ISOC, Villa{\~n}ueva de la Ca{\~n}ada, Madrid, Spain
           }

   \date{}

 
  \abstract{We present the results of the first four (quasi-)simultaneous radio (\atca), X-ray (\swift, \xte), and $\gamma$-ray (\integral) observations of \correc{the 
  black hole candidate}   \IGR, performed in February and March 2011.  \correc{The X-ray analysis shows that the source was in the hard state, and 
  then it transited 
  to a soft intermediate state. We study the correlated radio/X-ray behaviour of this source for the first time.}  
  The radio counterpart to \IGR\  was detected during 
   all four observations with the ATCA. In the hard state, the radio spectrum is typical of optically thick synchrotron emission from 
   a self-absorbed compact jet. In the soft intermediate state, the detection of optically thin synchrotron emission is probably 
   due to a discrete ejection event associated with the state transition.    
   \correc{The position of \IGR\  in the radio versus X-ray luminosity diagram (aka fundamental plane) is compatible with that of the other black hole sources
   for distances greater than 11 kpc. \IGR\  also appears as a new member of the  few 
 sources that show a strong quenching of radio emission after the state transition.} 
   Using the estimated luminosity at the spectral  transition from the hard state, and for a typical mass of $10 M_\odot$,  
   we estimate a distance to the source between $\sim$$11$ and $\sim$$17$ kpc, \correc{compatible with the radio behaviour of the source.  }}
 

   \maketitle
%

\section{Introduction}
Most \correc{of the $\sim$40 Galactic} black hole binaries (BHB) are  transient sources that are detected in  X-rays during bright, 
months-to-year long outbursts. 
During this period, their luminosity 
varies by several orders of magnitude. At the same time, the broad band ($\sim$$0.1$--$200$ keV) X-ray spectra show pivoting 
 from the canonical hard state (HS) to the canonical soft state (SS), through intermediate flavours of these last two. 
 These states also have clear signatures in the temporal  domain ($0.1$--$1000$ Hz), with the appearance of different
  types (dubbed A, B, C) of quasi periodic oscillations (QPO) in the HS, 
hard intermediate (HIMS), and soft intermediate (SIMS) states \citep[e.g.][]{Remillard06,homan05,remillard02}. \\
\indent BHBs are also known to produce strong radio emission, either from a powerful compact  jet in the HS or 
from relativistic discrete ejections at the transition from the HIMS to the SIMS \citep[e.g.,][]{fender06b}.  
In the HS, a strong correlation between the X-ray and radio luminosity 
has been found by  \citet{Corbel03} who point out a very strong coupling between the compact jets and the inner accretion flow. For Cyg X-1, this jet 
has been detected above $\sim$$500$~keV \citep{laurent11}. The \corr{$L_{\rm{Radio}}$=$L_{\rm{X}}^{0.6}$} relation between the radio and X-ray 
luminosities is thought to be universal \corr{(aka the ``standard" pattern or GX 339$-$4 like behaviour), although it  is} based on only 
eight Galactic sources  \citep[among which GX 339$-$4 and Cyg X-1 show a large quenching of the radio emission in the SS; ][]{Corbel03,fender10}. 
 \corr{Seven sources  have recently been shown to define a new track in the radio versus X-ray 
luminosity diagram by following  the relation $L_{\rm{Radio}}$=$L_{\rm{X}}^{1.4}$} 
\citep[\corr{hereafter the } H1743-322 like behaviour; ][]{fender10,mickael_1743}. 
The total number of BHBs simultaneously followed in a multi-wavelength way is, however, still low.Ê It is, thus,  not clear which 
of the two types of behaviour (GX 339$-$4 or H1743$-$322-like)
 is the most common, or what leads to these differences. Multi-wavelength observations of \correc{any} new active target, over the widest possible range of luminosities \correc{are, therefore, important for
  better understanding the jet behaviour and } assessing  the  links between the accretion and ejection processes.  \correc{To this end,  
we conducted Australian Telescope Compact Array (\atca) and Rossi X-ray Timing Explorer (\xte)Ê
observations as soon as \IGR\ was seen to enter a new outburst in 2011.}\\
\indent \IGR\ was discovered with the INTErnational Gamma-Ray Astrophysics Laboratory (\integral) on April 14, 2003 while 
entering  outburst \citep{2003ATel..149....1K}.
Observations with other facilities (e.g.  \xte, \swift) permitted  it to be classified as a low-mass 
X-ray binary, which was probably hosting a black hole \citep[e.g.][]{lutovinov05_5igr,capitanio09}. \correc{The renewal of activity  in  
late January 2011  \citep{krimm11a, krimm11b} } prompted \correc{several observing} campaigns on the source. 
It was, in particular, found that \IGR\ has similarities with GRS 1915+105, since it showed types of X-ray variability reminiscent of 
those of the so-called  ``$\rho$-heartbeat" and ``$\beta$" classes seen in GRS 1915+105  \citep{altamirano_b,altamirano_a}.
\correc{The specifics of the \IGR's X-ray behaviour further justify a (simultaneous) multi-wavelength study, to see, in particular,  
how it compares to the other $\sim$17 BHBs with known radio-X-ray behaviour.} \\
\indent Early in the outburst, we reported  the presence of radio emission compatible with a compact jet in the HS \citep{corbel11} 
and a $\sim$$0.1$~Hz QPO at the same time \citep{2011ATel.3168....1R}. Here, we present an in-depth analysis of 
  four (quasi) simultaneous radio-X-$\gamma$-ray observations. \correc{To complement the \atca\ and \xte\ data, we use } \integral\ and  \swift\  
\correc{observations to enhance the spectral coverage.}

\vspace*{-0.4cm}
\begin{table}[!h]
\centering
\caption{Radio fluxes of \IGR\ during the four \atca\ observations. Errors are at the 1$\sigma$ level.}
\begin{tabular}{lccc}
\hline
\hline
Obs. & $F_{5.5{\rm{GHz}}}$ & $F_{9{\rm{GHz}}}$ & $\alpha$\\
         & (mJy)                            &         (mJy)                 &                 \\
\hline
A1 & $1.40$$\pm$$0.05$ & $1.24$$\pm$$0.06$ & $-0.25$$\pm$$0.12$\\
A2 & $1.53$$\pm$$0.10$ & $1.57$$\pm$$0.10$ & $+0.05$$\pm$$0.19$\\
A3 & $2.41$$\pm$$0.10$ & $1.13$$\pm$$0.10$ & $-1.54$$\pm$$0.20$\\
A4 & $0.17$$\pm$$0.05$ & $<$$0.08$ & \\
\hline
\end{tabular}
\label{table:radio}
\end{table}
\vspace*{-0.4cm}

\section{Observations and data reduction}
\begin{figure}
\centering
\epsfig{file=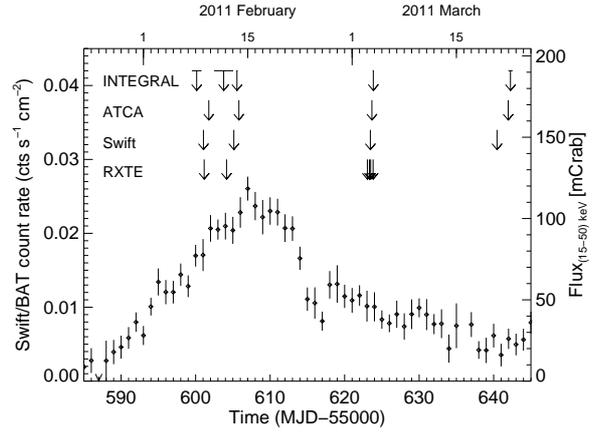,width=8.4 cm}
\caption{15--50 keV \swift/BAT lightcurve of \IGR. The vertical arrows represent the (mid) times of 
the pointed observations. The width of the arrow's top is the length of the observation (visible only for \integral).}
\label{fig:lc}
\end{figure}

Figure~\ref{fig:lc} shows the 15--50 keV daily averaged \swift/BAT light curve of \IGR, with the various
observations discussed in this paper. The  journal of these observations can be found \corr{online in Table~\ref{table:log}. 
The observations are, hereafter, ordered and labelled according to the instrument used (A stands for \atca, I for \integral, S for \swift, and R for \xte).} \
The \atca\ observations were made in two frequency bands (5.5 and 9 GHz) simultaneously,   
with the upgraded Compact Array Broadband Backend \citep{Wilson2011}. 
The amplitude and band-pass calibrator was PKS~1934$-$638, and the antenna's gain and phase calibration, 
as well as the polarisation leakage, were derived from regular observations 
of the nearby ($\sim$3.2$^\circ$away) calibrator PMN~1714$-$336.  See, e.g., \citet{Corbel04b} for more details on the standard data reduction.\\
\indent The \xte\ and \swift\ data were reduced with  
{\tt{HEASOFT\footnote{http://heasarc.gsfc.nasa.gov/docs/software/lheasoft/} v6.10}}, and  the \integral\ data  with 
 {\tt{OSA v9.0\footnote{http://www.isdc.unige.ch/integral/analysis\#Software}}}. \corr{The X-ray spectral fitting software package} {\tt{XSPEC 12.6.0q}} was 
 used to fit the energy and  power density spectra  (PDS) \corr{after converting  the  latter into {\tt{XSPEC}} readable files}.\\
\indent  The \swift/XRT level 2 cleaned event files 
were obtained from window timing (WT) mode data with {\tt{xrtpipeline}} keeping only  the grade 0 
events. We then extracted the spectra from a $40$$\times$$10$ pixel region centred on the 
best source position. The background was estimated from  a region of the same size at an off-axis 
position.  The ancillary response file (arf) was estimated with {\tt{xrtmkarf}}, and the last version (v. 12) of the WT 
redistribution matrix file (rmf) was used in the spectral fits.\\
\indent The \integral/IBIS/ISGRIÊ source spectra were obtained in a  standard way \citep[e.g.][]{rodrigue08_1915a}.  
We used the latest version of the rmf and arf files available 
from the \integral\ instrument calibration tree provided with {\tt{OSA 9.0}}. In two observations, the source is also within the 
field of view (fov) of the JEM-X monitors, and we extracted 16 channels spectra from JEM-X unit 2. \\
\indent The  Z source GX 349+2 is at  $\sim$$41$\arcmin\ from \IGR. The \xte/PCA data are thus contaminated by the emission 
from this bright source at a level that is  very difficult to estimate given the lack of strictly simultaneous data. 
The \xte/PCA data were therefore  only considered for timing analysis \citep[e.g.][]{2011ATel.3168....1R}. 
Two  to 60 keV PDS were extracted from {\tt{Good Xenon}}  mode date with {\tt{POWSPEC v1.0}} 
over the 0.0078125--500 Hz range.  \corr{The fits of these PDSs were 
restricted to the frequency range 0.0078125--30 Hz.} 

\section{Results}
\subsection{Radio observations}
The  analysis of the \atca\  data shows  there is a single radio source 
within the X-ray error circle, at a location of $(\alpha,\delta) = 17^{\rm{h}}$ $09^{\rm{m}}$ $07^{\rm{s}}\!.61$,  $-36^\circ$ $24\arcmin$ $25\arcsec\!.7$  
(J2000,  1$\sigma$ positional uncertainty of 0.1$\arcsec$). This position is also consistent with 
the position of the suggested optical/NIR counterpart \citep{torres11}. The measured radio fluxes and 
\correc{spectral} indices \correc{$\alpha$ (defined by F($\nu$)$\propto$$\nu^{\alpha}$)} are reported in Table~\ref{table:radio}.

\subsection{X and $\gamma$-ray spectral analysis}
\begin{figure}
\centering
\epsfig{file=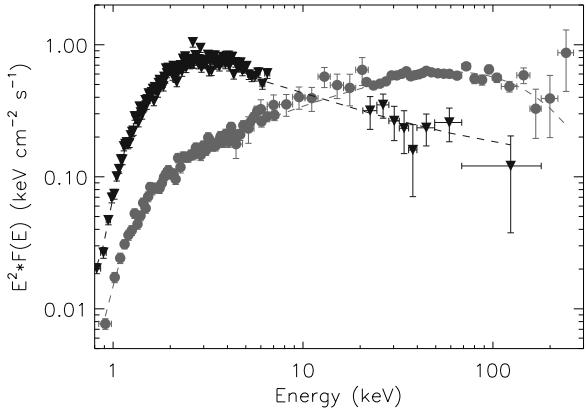,width=8.5cm}
\caption{Spectra of obs. S1+I1 (XRT/JEM-X 2/ISGRI, grey filled circles) and S3+I3 (XRT/ISGRI, dark triangles). 
 The dashed lines represent the best-fit  models.}
\label{fig:Spec}
\end{figure}

The  \swift\  and \integral\  spectra S1+I1, S2+I2, S3+I3, and S4+I4  were fitted simultaneously, and we retained the energy channels 
0.7--7.5~keV for XRT,  20--300 keV for ISGRI, and, in S1+I1 and I1', the  4--20 keV channels for JEM-X2.  
The best-fit models consist of an absorbed\footnote{Abundances fixed to those from \citet{wilms00}} power law with a high 
energy cut-off ({\tt{tbabs*cutoffpl}} in {\tt{XSPEC}}, hereafter CPL) for spectra S1+I1, I1', and S2+I2. A disc component is required in S3+I3 and S4+I4, 
while the cut-off is not needed any more ({\tt{tbabs*(diskbb+powerlaw)}}, hereafter DPL).  A normalisation constant was also included and 
frozen to 1 for the XRT spectra and left free to vary for the other instruments, except for the I1' fits, where it is frozen to 1 for ISGRI.
The best-fit parameters we obtained are reported in Table~\ref{table:specfit}, and the `$\nu$-$F_{\nu}$' spectra of S1+I1  and S3+I3  are 
represented in Fig.~\ref{fig:Spec}.

\begin{table*}[htbp]
\caption{Results of the spectral fits to the joint \swift/XRT and \integral/ISGRI (and JEM-X in I1 and I1') spectra. Errors are at the 90\% level.}
\begin{tabular}{lllccccccc}
\hline
\hline
Spectra & model & \nh & kT$_{bb}$/kT$_{inj}$ & $\Gamma$/$\tau$ & E$_{\rm cut}$/ kT$_e$& \chisq & \multicolumn{3}{c}{Unabs. Flux$^\star$} \\
                     &             & ($\times 10^{22}$~cm$^{-2}$) & (keV) &  & (keV) &      (dof)   &  0.5--10 keV & 3--9 keV & 20--200 keV \\
\hline
S1+I1 & CPL & $1.08\pm0.08$ & & $1.40\pm0.07$ & $100_{-13}^{+17}$ & 0.90 (89) & 8.4 & 4.5 & 20.0\\
 
            & ThC & $1.14_{-0.05}^{+0.06}$ & 0.2 (frozen) & $0.8\pm0.3$ & $85_{-22}^{+43}$  & 1.00 (89) & 8.5 & 4.3 & 17.2 \\

I1'        & CPL                     & 1.1 frozen & & $1.49\pm0.09$ & $91_{-13}^{+17}$ & 1.0 (34) & NA & 6.5 & 22.4\\
            & ThC                     & 1.1 frozen & 0.2 (frozen) &  $0.8\pm0.2$ & $74_{-10}^{+22}$ & 1.0 (34) &NA & 7.2 & 22.3 \\

S2+I2 & CPL                      & $1.15\pm0.06$ &  & $1.54\pm0.05$ & $107_{-32}^{+61}$ & 1.1 (137) & 12.1 & 6.0 & 20.8\\
            & ThC                      & $1.13\pm0.05$ &0.2 (frozen) & $2.1_{-1.5}^{+0.5}$ & $27_{-7}^{+74}$& 1.2 (137) & 11.7 & 6.0 & 25.9 \\
            
S3+I3 & DPL & $1.5\pm0.1$ & $1.2\pm0.2$ & $2.3\pm0.3$  & & 1.3 (128) & 38.2 &  11.4 & 7.3 \\

S4+I4$^\circ$ &  DPL & $1.16\pm0.05$ & $1.29_{-0.03}^{+0.01}$ & $2.1_{-0.3}^{+0.1}$ & &1.2 (409)  & 31.6 &  12.4  & 2.1 \\
\hline
\hline
\end{tabular}
\begin{list}{}{}
\item[$^\star$]In units of $10^{-10}$~\ergcms. Fluxes are normalised to \swift/XRT, except  I1' normalised to ISGRI
\item[$^\circ$]Observations made more than a day apart. Norm. constant is high  and poorly constrained ($\gtrsim2$). 
\end{list}
\label{table:specfit}
\end{table*}

In XRBs, a cut-off power-law spectrum is usually interpreted as the signature of inverse  
Comptonisation of  soft seed photons by a thermalised (i.e., with velocities following a Maxwellian distribution) 
population of electrons. A thermal Comptonisation model \citep[{\tt{comptt}}][]{titarchuk94} provided  
 a good fit to  S1+I1, I1', and S2+I2 (ThC in Table~\ref{table:specfit}). We assumed a disc geometry in {\tt{XSPEC}} and  fixed the temperature 
 of the seed photons.   This model also permits  avoiding the divergence 
of the power-law flux towards low energy
 since Comptonisation significantly  contributes above $3$$\times$$kT_{inj}$ and decreases quickly below. 

\subsection{X-ray timing analysis}
\begin{figure}[htpb]
\centering
\epsfig{file=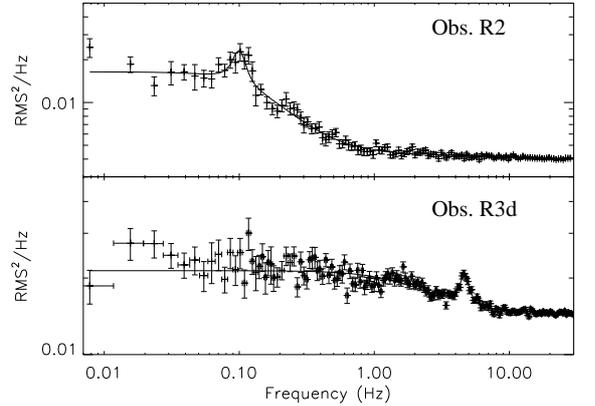,width=7.5cm}
\caption{PDSs of observations R2 (top panel) and R3d (bottom panel).}
\label{fig:pds}
\end{figure}

None of the \xte\ observations shows the  ``$\rho$-heartbeat" type of variability. 
Since GX 349+2 contaminates significantly Obs. R1 and R2 \correc{(Sec. 2)}, we do not discuss the 
level of continuum variability since we cannot  attribute it to \IGR\ with certainty. The \IGR\  PDSs, however, show 
QPOs that are not associated with GX 349+2 \citep{2011ATel.3168....1R,shapo11,pahari11}. To obtain their `true' 
amplitudes during Obs. R1 and R2, we have estimated the net count rate of \IGR\ in the \xte/PCA by using 
our spectral results (Table~\ref{table:specfit}) and simulated a PCA 
spectrum in {\tt{XSPEC}}. The amplitude of the QPO is then  
$A_{\rm{net}}$=$A_{\rm{raw}}\times\frac{CR_{\rm{PCA}}}{CR_{\rm{net}}}$, where $A_{\rm{raw}}$ is the rms amplitude of the QPO obtained from the fit 
to the PDS, $CR_{\rm{net}}$  the simulated count rate of \IGR, and $CR_{\rm{PCA}}$  the observed PCA rate. 
No correction was applied to R3a,b,c,d since GX 349+2 was out of the PCA fov. The  QPO parameters \corr{obtained from the analysis} 
are reported in Table~\ref{tab:qpo}, and two representative PDSs are plotted in Fig.~\ref{fig:pds}.
These QPOs can be classified as type C QPOs, except maybe those found in observations R3 that could be transitory between C and B.

\begin{table}[!h]
\centering
\caption{Parameters of the QPOs seen during Obs. R1, R2, and R3. Errors are given at the 90\% confidence level.}
\begin{tabular}{lccc}
\hline
\hline
PDS & $\nu_{\rm{QPO}}$ & Q & A\\
          &  (Hz)                        &      (=$\nu/FWHM$)                         & (\% rms)\\
\hline
R1    &   $0.081_{-0.004}^{+0.005}$ & 4.3$\pm2.2$ & $12.3_{-1.7}^{+1.8}$\\
R2    & $0.100_{-0.005}^{+0.008}$ & 4.2$\pm1.5$ & $10.1_{-1.6}^{+1.2}$ \\
R3a  & $4.57\pm0.08$             & $4.2\pm1.1$ & $8.8\pm1.2$\\
R3b  & $4.57\pm0.08$             & $4.5\pm1.3$ & $8.5_{-1.3}^{+1.1}$\\
R3c & $4.28\pm0.06$              &  $2.4\pm0.3$ & $12.5_{-0.7}^{+0.5}$\\
R3d & $4.66\pm0.05$              & $5.8\pm1.5$ & $8.0\pm0.6$\\
\hline
\end{tabular}
\label{tab:qpo}
\end{table}

\vspace*{-0.5cm}
\vspace*{-0.7cm}
\section{Discussion}
\begin{figure}[htpb]
\epsfig{file=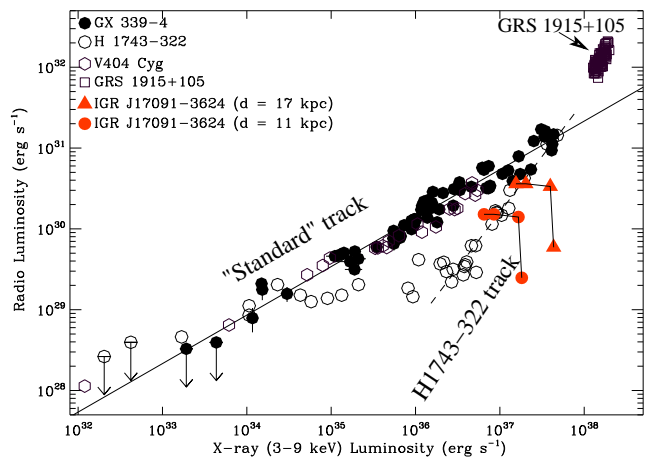,width=9cm}
\caption{Radio vs X-ray luminosity for several BHBs.}
\label{fig:radioX}
\end{figure}

In the first two observations, the spectral parameters  
(Table~\ref{table:specfit}) and presence of type C QPOs  are typical of a BH in the HS \citep[e.g.][]{Remillard06, homan05}.  During the 
third observation, the source clearly is in a softer state. The transition is associated with a discrete radio ejection (see below) 
\correc{as often seen in BHBs \citep[the clearest example being GRS 1915+105, e.g.][]{rodrigue08_1915a}}. This, 
 and the presence of QPOs at  4--5 Hz,  indicates that \IGR\ is in a  SIMS at that time. 
 No temporal (\xte) data  are available for the fourth observation. The spectra are indicative of either a  SIMS or a pure SS.  \\
\indent The value of \nh\  obtained from the spectral fits is slightly higher than the $2^\circ$$\times$$2^\circ$ 
weighted averaged absorption in this direction  \citep[\nh=$6.0$$\times$$10^{21}$~cm$^{-2}$; ][]{kalberla05}. 
This is consistent with a  source lying at a rather large distance in the Galaxy.\\
\indent The HS$\rightarrow$SIMS state transition occurred between MJDs 55612 and 55616 \citep{pahari11}. Obs. 2 is the observation in the HS 
that is the closest to the peak of the hard X-ray outburst (Fig.~Ê\ref{fig:lc}) and to the state transition. It has  a bolometric 
flux of F$_{\rm{bol, trans}}$$\sim$$4.1$$\times$$10^{-9}$~\ergcms\ (ThC model).   
Most BHBs have $L_{\rm{trans,  HS\rightarrow SS}}$$\gtrsim$$4\%$$L_{\rm{Edd, 1-200keV}}$, when assuming 
a power-law spectrum with $\Gamma$$=$$2$ \citep{yu09}\footnote{With exception of  Cyg X-1 (HMXB) and 
GRO J1655$-$40 \citep[discussed as an outlier by][]{yu09}.}. Although this value slightly underestimates
 the bolometric luminosity at the transition, we take it as a `hard' lower limit 
on $L_{\rm{trans}}$, which  therefore allows us to estimate a lower limit on the distance to the source. 
Using F$_{\rm{bol, trans}}$ estimated above, we obtain a 
distance $d_{J17091}$$\sim$$11$~kpc for a BH of  $\sim$10$M_\odot$. 
With $L_{\rm{trans}}$$=$$10\% L_{\rm{Edd}}$ \citep{esin97} we obtain $d_{J17091}$$\sim$$17$ kpc\footnote{Fit of  non simultaneous
\swift/\integral\ observations performed on MJDs 55611--55612 leads to  F$_{\rm{bol, trans}}$$\sim$$3.5$$\times$$10^{-9}$\ergcms. This lower 
value would place \IGR\ at an even farther distance.}. Using a \correc{different method (based on the QPO frequency-photon index correlation)}, 
\citet{pahari11}  also favour  a large distance  with a range of masses between 8--11.4$M_\odot$.\\
\indent  The radio spectra
of observations \correc{A1 and A2} are consistent with being flat and they indicate of the presence of self-absorbed 
compact jets (e.g. \citealt{Corbel04b, fender04b}). \correc{Observations A3 and A4} are  characteristic of optically thin synchrotron emission 
and can be attributed to the emission of a discrete ejection \correc{associated with} the hard to soft  transition.  \correc{The observation
of yet another BHB showing discrete ejections at  the hard to soft transition probably indicates a link between 
transitions and (discrete) ejections potentially common to all microquasars. In XTE J1550$-$564 and GRS 1915+105, 
\citet{rodrigue03_1550, rodrigue08_1915a, rodrigue08_1915b} have suggested that the Comptonised component was the source of 
the ejected material. Here, the change in spectral shape and drop in the hard X-ray flux (Table \ref{table:specfit}) is compatible with this interpretation.}  \\
\indent  In Fig.~\ref{fig:radioX}, we plot  the levels of radio  and X-ray emission for  \IGR\ in comparison with a sample of representative 
BHBs \citep[data from][]{Corbel03,Corbel08,mickael_1743,Rushton10} assuming the distances 
estimated above. The two radio points with the lowest X-ray luminosities correspond to the HS, while the other two correspond to 
the optically thin radio flare. At a distance of 11 kpc, the source behaviour joins the track 
followed by  H1743$-$322 (Fig. \ref{fig:radioX}). In this case \corr{(and at any lower distance)}, however, the \corr{break of the radio vs X-ray relation occurs 
at  too low luminosities} compared to the standard BHBs \corr{\citep{fender10}}. 
 At 17 kpc  the  position of \IGR\  is consistent with the radio/X-ray correlation of most BHBs (within the 
  dispersion), although we cannot  assign it precisely to a specific track (i.e. the GX~339$-$4 or the H~1743$-$322 one, Fig.~\ref{fig:radioX}). In any cases, a 
  large distance ($\gtrsim$ 11 kpc) to the source is favoured by the radio-X-ray behaviour.\\
\indent \correc{\IGR\ has  recently been compared  to GRS 1915+105 \citep{altamirano_b,altamirano_a,pahari11}.  
However, even at the favoured distance of 17 kpc,  our simultaneous radio-X-ray observations show that \IGR\  does not reach the bright levels of radio luminosity observed for GRS~1915$+$105 (Fig. \ref{fig:radioX}), and is more similar to standard BHBs, such as  
GX 339$-$4 or H1743$-$322. This would tend to indicate that the radio properties discussed here are not related  to the fast flaring X-ray behaviour 
common to both GRS 1915+105 and \IGR.}

\vspace*{-0.3cm}

\begin{acknowledgements}
We thank P. Varni\`ere, R. Owen, and P. Curran for useful discussions. JR and SC acknowledge 
partial funding from the European FP7  grant 
agreement number ITN 215212 ``Black Hole Universe". IC is supported by the CNES through CNRS. 
JAT acknowledges partial support from NASA via INTEGRAL
and Swift Guest Observer grants NNX08AX93G and NNX08AW35G. AP acknowledges financial contribution from the agreement ASI-INAF I/009/10/0.
The Australia Telescope is funded by the Commonwealth of Australia for operation as a national
facility managed by CSIRO. We acknowledge the use of  the IGR Sources page (http://irfu.cea.fr/Sap/IGR-Sources/), the use of public  \swift\ and \xte\ data obtained through the HEASARC Online Service, 
provided by  NASA/GSFC, and observations made with
 \integral, an ESA project with instruments and science data centre funded by ESA member states, Poland and with the participation 
 of Russia and the USA.  
\end{acknowledgements}
\bibliographystyle{aa}

\begin{thebibliography}{34}
\expandafter\ifx\csname natexlab\endcsname\relax\def\natexlab#1{#1}\fi

\bibitem[{{Altamirano} {et~al.}(2011{\natexlab{a}}){Altamirano}, {Belloni},
  {Krimm}, {Casella}, {Curran}, {Kennea}, {Kalamkar}, {van der Klis},
  {Wijnands}, {Linares}, {Motta}, {Munoz-Darias}, \& {Stiele}}]{altamirano_b}
{Altamirano}, D.,  {et~al.} 2011{\natexlab{a}}, The
  Astronomer's Telegram, 3299, 1

\bibitem[{{Altamirano} {et~al.}(2011{\natexlab{b}}){Altamirano}, {Linares},
  {van der Klis}, {Wijnands}, {Kalamkar}, {Casella}, {Watts}, {Patruno},
  {Armas-Padilla}, {Cavecchi}, {Degenaar}, {Kaur}, {Yang}, \&
  {Rea}}]{altamirano_a}
{Altamirano}, D., {et~al.}
  2011{\natexlab{b}}, The Astronomer's Telegram, 3225, 1

\bibitem[{{Capitanio} {et~al.}(2009){Capitanio}, {Giroletti}, {Molina},
  {Bazzano}, {Tarana}, {Kennea}, {Dean}, {Hill}, {Tavani}, \&
  {Ubertini}}]{capitanio09}
{Capitanio}, F., {Giroletti}, M., {Molina}, M., {et~al.} 2009, \apj, 690, 1621

\bibitem[{{Corbel} {et~al.}(2004){Corbel}, {Fender}, {Tomsick}, {Tzioumis}, \&
  {Tingay}}]{Corbel04b}
{Corbel}, S., {Fender}, R.~P., {Tomsick}, J.~A., {Tzioumis}, A.~K., \&
  {Tingay}, S. 2004, \apj, 617, 1272

\bibitem[{{Corbel} {et~al.}(2008){Corbel}, {Koerding}, \& {Kaaret}}]{Corbel08}
{Corbel}, S., {Koerding}, E., \& {Kaaret}, P. 2008, \mnras, 389, 1697

\bibitem[{{Corbel} {et~al.}(2003){Corbel}, {Nowak}, {Fender}, {Tzioumis}, \&
  {Markoff}}]{Corbel03}
{Corbel}, S., {Nowak}, M.~A., {Fender}, R.~P., {Tzioumis}, A.~K., \& {Markoff},
  S. 2003, \aap, 400, 1007

\bibitem[{{Corbel} {et~al.}(2011){Corbel}, {Rodriguez}, {Tzioumis}, \&
  {Tomsick}}]{corbel11}
{Corbel}, S., {et~al.} 2011, The
  Astronomer's Telegram, 3167, 1

\bibitem[{{Coriat} {et~al.}(2011){Coriat}, {Corbel}, {Prat}, {Miller-Jones},
  {Cseh}, {Tzioumis}, {Brocksopp}, {Rodriguez}, {Fender}, \&
  {Sivakoff}}]{mickael_1743}
{Coriat}, M., {Corbel}, S., {Prat}, L., {et~al.} 2011, \mnras, 619

\bibitem[{{Esin} {et~al.}(1997){Esin}, {McClintock}, \& {Narayan}}]{esin97}
{Esin}, A.~A., {McClintock}, J.~E., \& {Narayan}, R. 1997, \apj, 489, 865

\bibitem[{{Fender}(2006)}]{fender06b}
{Fender}, R. 2006, {Jets from X-ray binaries} (Compact stellar X-ray sources),
  381--419

\bibitem[{{Fender} {et~al.}(2004){Fender}, {Belloni}, \& {Gallo}}]{fender04b}
{Fender}, R.~P., {Belloni}, T.~M., \& {Gallo}, E. 2004, \mnras, 355, 1105

\bibitem[{{Fender} {et~al.}(2010){Fender}, {Gallo}, \& {Russell}}]{fender10}
{Fender}, R.~P., {Gallo}, E., \& {Russell}, D. 2010, \mnras, 406, 1425

\bibitem[{{Homan} {et~al.}(2005){Homan}, {Buxton}, {Markoff}, {Bailyn},
  {Nespoli}, \& {Belloni}}]{homan05}
{Homan}, J., {Buxton}, M., {Markoff}, S., {et~al.} 2005, \apj, 624, 295

\bibitem[{{Kalberla} {et~al.}(2005){Kalberla}, {Burton}, {Hartmann}, {Arnal},
  {Bajaja}, {Morras}, \& {P{\"o}ppel}}]{kalberla05}
{Kalberla}, P.~M.~W., {Burton}, W.~B., {Hartmann}, D., {et~al.} 2005, \aap,
  440, 775

\bibitem[{{Krimm} {et~al.}(2011){Krimm}, {Barthelmy}, {Baumgartner},
  {Cummings}, {Fenimore}, {Gehrels}, {Kennea}, {Markwardt}, {Palmer},
  {Sakamoto}, {Skinner}, {Stamatikos}, {Tueller}, \& {Ukwatta}}]{krimm11a}
{Krimm}, H.~A., {et~al.} 2011, The
  Astronomer's Telegram, 3144, 1

\bibitem[{{Krimm} \& {Kennea}(2011)}]{krimm11b}
{Krimm}, H.~A. \& {Kennea}, J.~A. 2011, The Astronomer's Telegram, 3148, 1

\bibitem[{{Kuulkers} {et~al.}(2003){Kuulkers}, {Lutovinov}, {Parmar},
  {Capitanio}, {Mowlavi}, \& {Hermsen}}]{2003ATel..149....1K}
{Kuulkers}, E.,  {et~al.} 2003, The Astronomer's
  Telegram, 149, 1

\bibitem[{{Laurent} {et~al.}(2011){Laurent}, {Rodriguez}, {Wilms}, {Cadolle
  Bel}, {Pottschmidt}, \& {Grinberg}}]{laurent11}
{Laurent}, P., {Rodriguez}, J., {Wilms}, J., {et~al.} 2011, Science, 332, 438

\bibitem[{{Lutovinov} {et~al.}(2005){Lutovinov}, {Revnivtsev}, {Molkov}, \&
  {Sunyaev}}]{lutovinov05_5igr}
{Lutovinov}, A., {Revnivtsev}, M., {Molkov}, S., \& {Sunyaev}, R. 2005, \aap,
  430, 997

\bibitem[{{Pahari} {et~al.}(2011){Pahari}, {Yadav}, \&
  {Bhattacharyya}}]{pahari11}
{Pahari}, M., {Yadav}, J., \& {Bhattacharyya}, S. 2011, submitted, ArXiv 1105.4694

\bibitem[{{Remillard} \& {McClintock}(2006)}]{Remillard06}
{Remillard}, R.~A. \& {McClintock}, J.~E. 2006, \araa, 44, 49

\bibitem[{{Remillard} {et~al.}(2002){Remillard}, {Sobczak}, {Muno}, \&
  {McClintock}}]{remillard02}
{Remillard}, R.~A., {Sobczak}, G.~J., {Muno}, M.~P., \& {McClintock}, J.~E.
  2002, \apj, 564, 962

\bibitem[{{Rodriguez} {et~al.}(2003){Rodriguez}, {Corbel}, \&
  {Tomsick}}]{rodrigue03_1550}
{Rodriguez}, J., {Corbel}, S., \& {Tomsick}, J.~A. 2003, \apj, 595, 1032

\bibitem[{{Rodriguez} {et~al.}(2011){Rodriguez}, {Corbel}, {Tomsick}, {Paizis},
  \& {Kuulkers}}]{2011ATel.3168....1R}
{Rodriguez}, J.,  {et~al.} 2011, The Astronomer's Telegram, 3168, 1

\bibitem[{{Rodriguez} {et~al.}(2008{\natexlab{a}}){Rodriguez}, {Hannikainen},
  {Shaw}, {Pooley}, {Corbel}, {Tagger}, {Mirabel}, {Belloni}, {Cabanac},
  {Cadolle Bel}, {Chenevez}, {Kretschmar}, {Lehto}, {Paizis}, {Varni{\`e}re},
  \& {Vilhu}}]{rodrigue08_1915a}
{Rodriguez}, J., {Hannikainen}, D.~C., {Shaw}, S.~E., {et~al.}
  2008{\natexlab{a}}, \apj, 675, 1436

\bibitem[{{Rodriguez} {et~al.}(2008{\natexlab{b}}){Rodriguez}, {Shaw},
  {Hannikainen}, {Belloni}, {Corbel}, {Cadolle Bel}, {Chenevez}, {Prat},
  {Kretschmar}, {Lehto}, {Mirabel}, {Paizis}, {Pooley}, {Tagger},
  {Varni{\`e}re}, {Cabanac}, \& {Vilhu}}]{rodrigue08_1915b}
{Rodriguez}, J., {Shaw}, S.~E., {Hannikainen}, D.~C., {et~al.}
  2008{\natexlab{b}}, \apj, 675, 1449

\bibitem[{{Rushton} {et~al.}(2010){Rushton}, {Spencer}, {Fender}, \&
  {Pooley}}]{Rushton10}
{Rushton}, A., {Spencer}, R., {Fender}, R., \& {Pooley}, G. 2010, \aap, 524,
  A29+

\bibitem[{{Shaposhnikov}(2011)}]{shapo11}
{Shaposhnikov}, N. 2011, The Astronomer's Telegram, 3179, 1

\bibitem[{{Titarchuk}(1994)}]{titarchuk94}
{Titarchuk}, L. 1994, \apj, 434, 570

\bibitem[{{Torres} {et~al.}(2011){Torres}, {Jonker}, {Steeghs}, \&
  {Mulchaey}}]{torres11}
{Torres}, M.~A.~P., , {et~al.}  2011,
  The Astronomer's Telegram, 3150, 1

\bibitem[{{Wilms} {et~al.}(2000){Wilms}, {Allen}, \& {McCray}}]{wilms00}
{Wilms}, J., {Allen}, A., \& {McCray}, R. 2000, \apj, 542, 914

\bibitem[{{Wilson} {et~al.}(2011){Wilson}, {Ferris}, {Axtens}, {Brown},
  {Davis}, {Hampson}, {Leach}, {Roberts}, {Saunders}, {Koribalski}, {Caswell},
  {Lenc}, {Stevens}, {Voronkov}, {Wieringa}, {Brooks}, {Edwards}, {Ekers},
  {Emonts}, {Hindson}, {Johnston}, {Maddison}, {Mahony}, {Malu}, {Massardi},
  {Mao}, {McConnell}, {Norris}, {Schnitzeler}, {Subrahmanyan}, {Urquhart},
  {Thompson}, \& {Wark}}]{Wilson2011}
{Wilson}, W.~E., {Ferris}, R.~H., {Axtens}, P., {et~al.} 2011, ArXiv 1105.3532

\bibitem[{{Yu} \& {Yan}(2009)}]{yu09}
{Yu}, W. \& {Yan}, Z. 2009, \apj, 701, 1940

\end{thebibliography}

\onltab{4}{
\begin{table}[htbp]
\caption{Journal of the multi-wavelength observations of \IGR\ presented in this paper.}
\begin{tabular}{llccc}
\hline
\hline
Facility & label & Obs. Id & MJD start &  Good time\\
              &           & (when available) & (d) &  \\
\hline
\atca & A1 &  & 55601.64 & 3.1 h \\
            & A2 &  & 55605.73 & 1.9 h\\
            & A3 &  & 55623.57 & 2.2 h\\
            & A4 &  & 55641.85 &  2.4 h \\
\xte & R1 &96103-01-01-00 & 55601.15 & 2486 s\\
       & R2 &96103-01-02-00 & 55604.15 & 5668 s \\
       & R3a &96420-01-01-010 &55623.05 & 6755 s \\
       & R3b & 96420-01-01-01 &  55623.32 & 4695 s\\
       & R3c &96420-01-01-020 & 55623.49 & 16263 s \\
       & R3d &96420-01-01-02 & 55623.82  & 7834 s\\
\swift & S1 & 00031921005 & 55601.05  & 1457 s \\
          & S2 & 00031921009 & 55605.13 & 2167 s \\
          & S3 & 00031921019 & 55623.45 & 686 s\\
          & S4 & 00031921030 & 55640.45 & 2352 s\\
\integral & I1 & Rev 1016$^\star$ & 55599.49 & 61080 s\\
                & I1'$^\dagger$ & Rev 1017 & 55602.49 & 131000 s\\
                & I2 & Rev 1018 & 55605.48 & 7818 s\\
                & I3 & Rev 1024 & 55623.77 & 6417 s\\
                & I4 & Rev 1030 & 55641.98 & 8583 s\\

\hline
\hline
\end{tabular}
\begin{list}{}{}
\item[$^\star$]\integral\ revolution number.
\item[$^\dagger$]I1' is intermediate (in time) between Obs. I1 and Obs. I2.
\end{list}
\label{table:log}
\end{table}
}
\end{document}